\documentclass[osajnl,twocolumn,show pacs,superscriptaddress,10pt]{revtex4-1} %% use 11pt for Applied Optics
\usepackage{amsmath,amssymb,graphicx}
\usepackage{color}
\usepackage{soul}
 \graphicspath{ {./} }

\begin{document}

\title{Coherent control of light interaction with graphene}

\author{Shraddha M Rao}
\author{Julius J F Heitz}
\author{Thomas Roger}
\author{Niclas Westerberg}
\author{Daniele Faccio}\email{Corresponding author: d.faccio@hw.ac.uk}
\affiliation{School of Engineering and Physical Sciences, Heriot-Watt University, Riccarton, Edinburgh, UK}

\begin{abstract} We report the experimental observation of all-optical modulation of light in a graphene film. The graphene film is scanned across a standing wave formed by two counter-propagating laser beams in a Sagnac interferometer. Through a coherent absorption process the on-axis transmission is modulated with close to 80\% efficiency. Furthermore we observe modulation of the scattered energy by mapping the off-axis scattered optical signal: scattering is minimized at a node of the standing wave pattern and maximized at an antinode. The results highlight the possibility to switch and modulate any given optical interaction with deeply sub-wavelength films.

\end{abstract}


\maketitle

\begin{figure}[!t]
  \includegraphics[width=8cm]{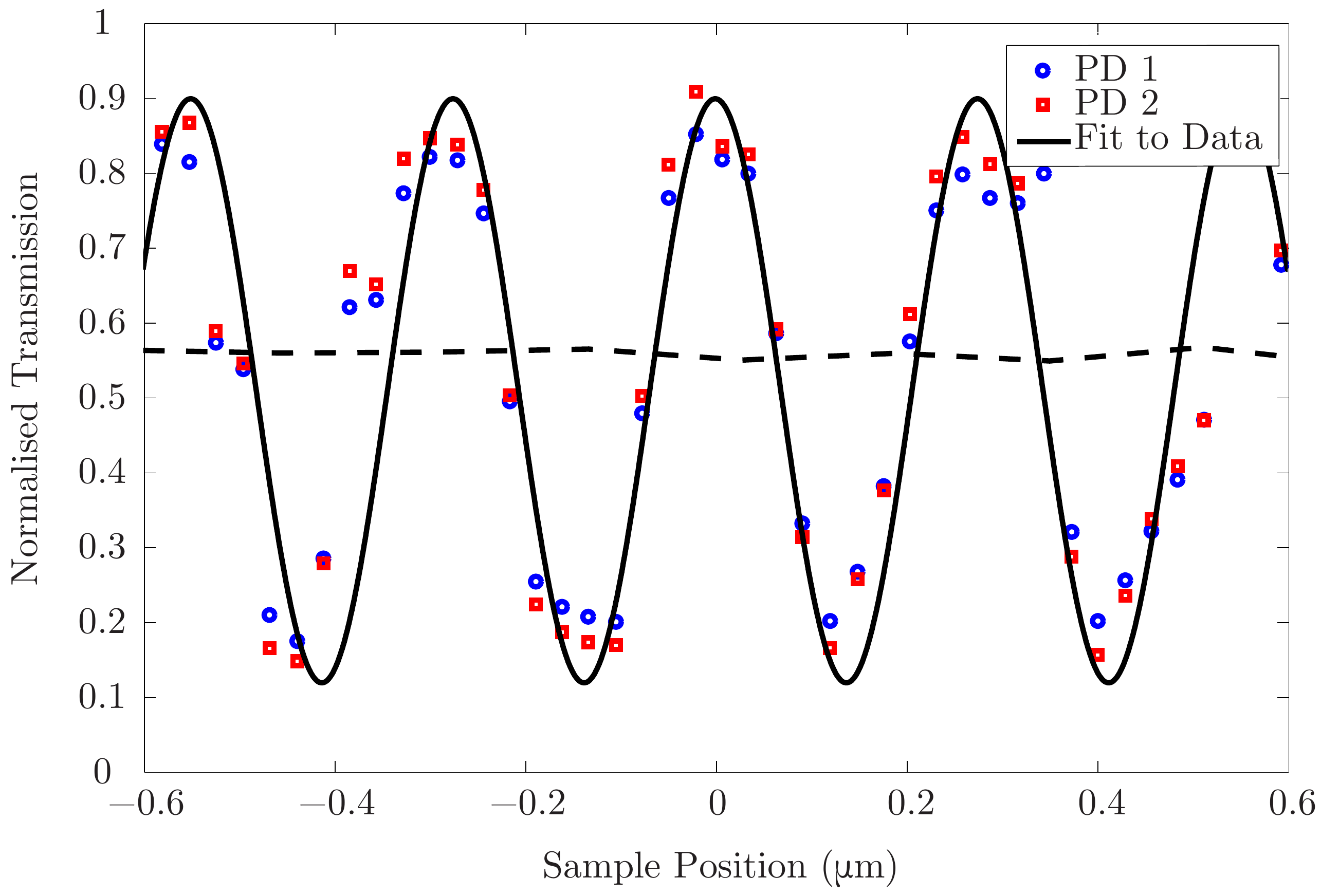}
\caption{Experimental data: Coherent absorption with CW, 532 nm illumination, from a multilayer graphene film, as a function of the sample position within the standing wave. The dashed line represents the absorption of the graphene sample.}
\label{fig:CW}
\end{figure}

\begin{figure}[!t]
  \includegraphics[width=8cm]{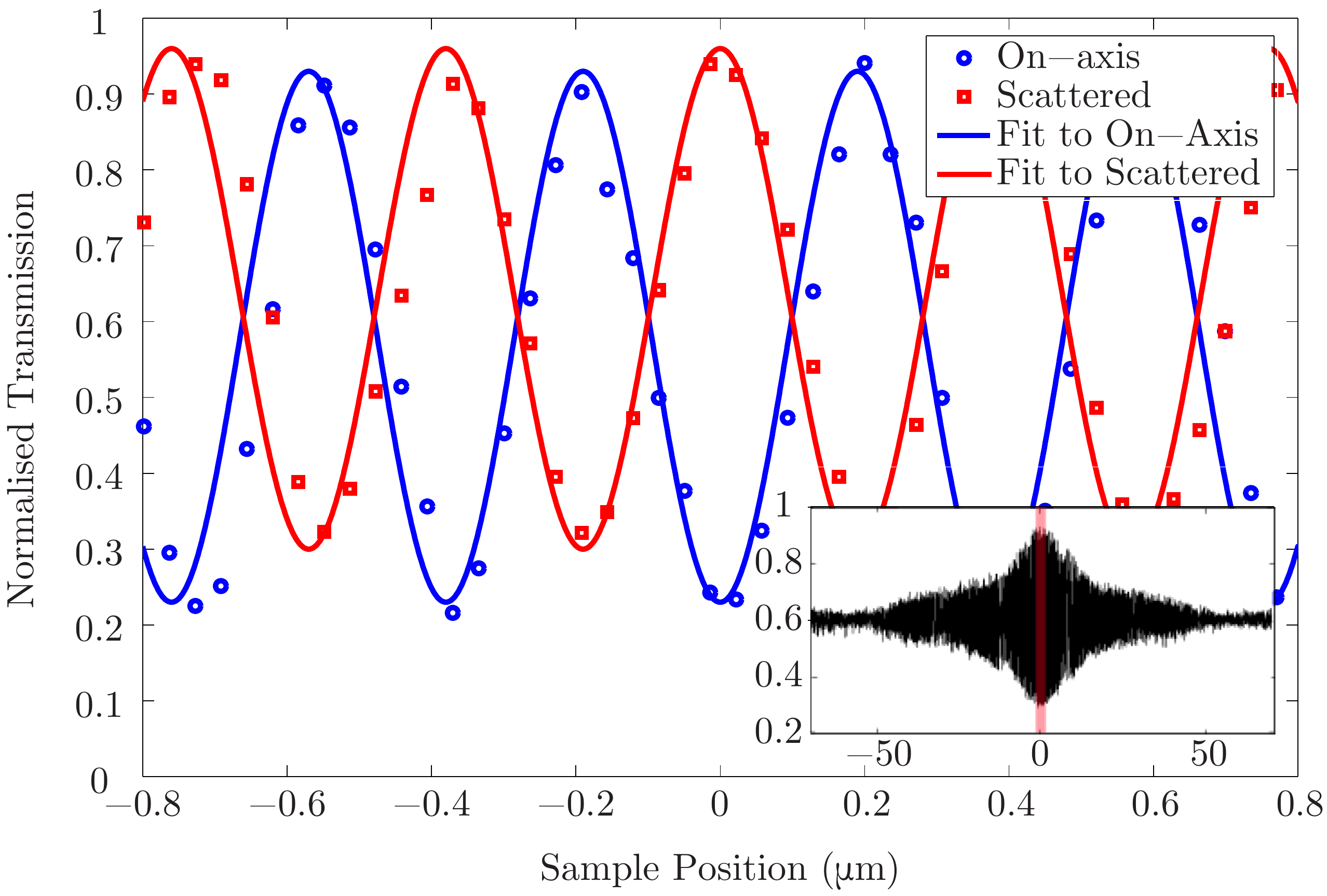}
\caption{Experimental data: Modulation of the transmitted light energy at 785 nm, using a pulsed laser source. On-axis transmission modulation (blue circles) and scattered energy (red squares) as a function of the sample position within the standing wave. Both data are fit with sine functions oscillating at twice the laser frequency. The inset shows the full autocorrelation trace.}
\label{fig:modulation}
\end{figure}

Sub-wavelength media, such as metamaterials, graphene and heterostructure layered media \cite{review0,review1,review2}, have great potential to provide enhanced functionality to optomechanical and photonic devices. A major limitation in the use of these materials is inefficient light-matter coupling at the nanoscale. Here, we examine the possibility to control both the absorption and scattering of light from graphene films through a process related to `coherent perfect absorption' (CPA).\\
\indent Coherent perfect absorption was first proposed in thick absorbing media, i.e., thicker than the optical wavelength \cite{Stone,Cao}, whereby the medium acts as a resonator cavity. Through a mechanism similar to a time-reversed laser \cite{t-reverse1, t-reverse2}, two coherent, monochromatic counterpropagating beams will be perfectly absorbed. These processes have been further developed in alternative geometries \cite{Agarwal, pirruccio}, including coherent absorption in nonlinear media \cite{Longhi}. \\
\indent It has recently been shown that it is also possible to observe CPA in a sub-wavelength film, allowing the coherent modulation of `light-with-light' without the use of nonlinearity \cite{zheludev,Zheludev1,Zheludev2}. This is possible due to the intricate nature of light-matter interaction at the nanoscale. Coherent light-with-light modulation in sub-wavelength films may arise in the presence of two counter-propagating light beams, \emph{a} and \emph{b}, that form a standing wave (within the coherence length of the laser). A sub-wavelength film that is placed in the path of the counter-propagating beams experiences a very different electric field amplitude, depending on its position within the standing wave. If the film is positioned within a node of the standing wave, the electric field amplitude is constant in time and equal to zero: the light beams will pass the film without loss. However at an anti-node, the position of maximum amplitude of the field, the light beams undergo strong, ideally 100\% absorption. This provides a method to modulate the absorption of light with a thin absorbing film by simply changing the relative phases of the interacting beams. This mechanism was first reported using a metamaterial thin film structure with properties tuned so as to provide a single-pass absorption close to the ideal 50\% limit that is required for perfect 100\% coherent absorption \cite{zheludev,abajo}. In single layer graphene, similar absorption properties have been proposed by patterning of the graphene layer \cite{abajo,theory} or by using geometries for increased light interaction \cite{modulator,Ferreira}. Similar to metallic metamaterial structures \cite{Qiu,Baumberg,Hagglund,Giessen}, structured graphene relies on a resonance between the light field and the structure itself, thus limiting operation to a specific wavelength range \cite{abajo,Yang}. \\
\begin{figure}[!t]
 \includegraphics[width = 8cm]{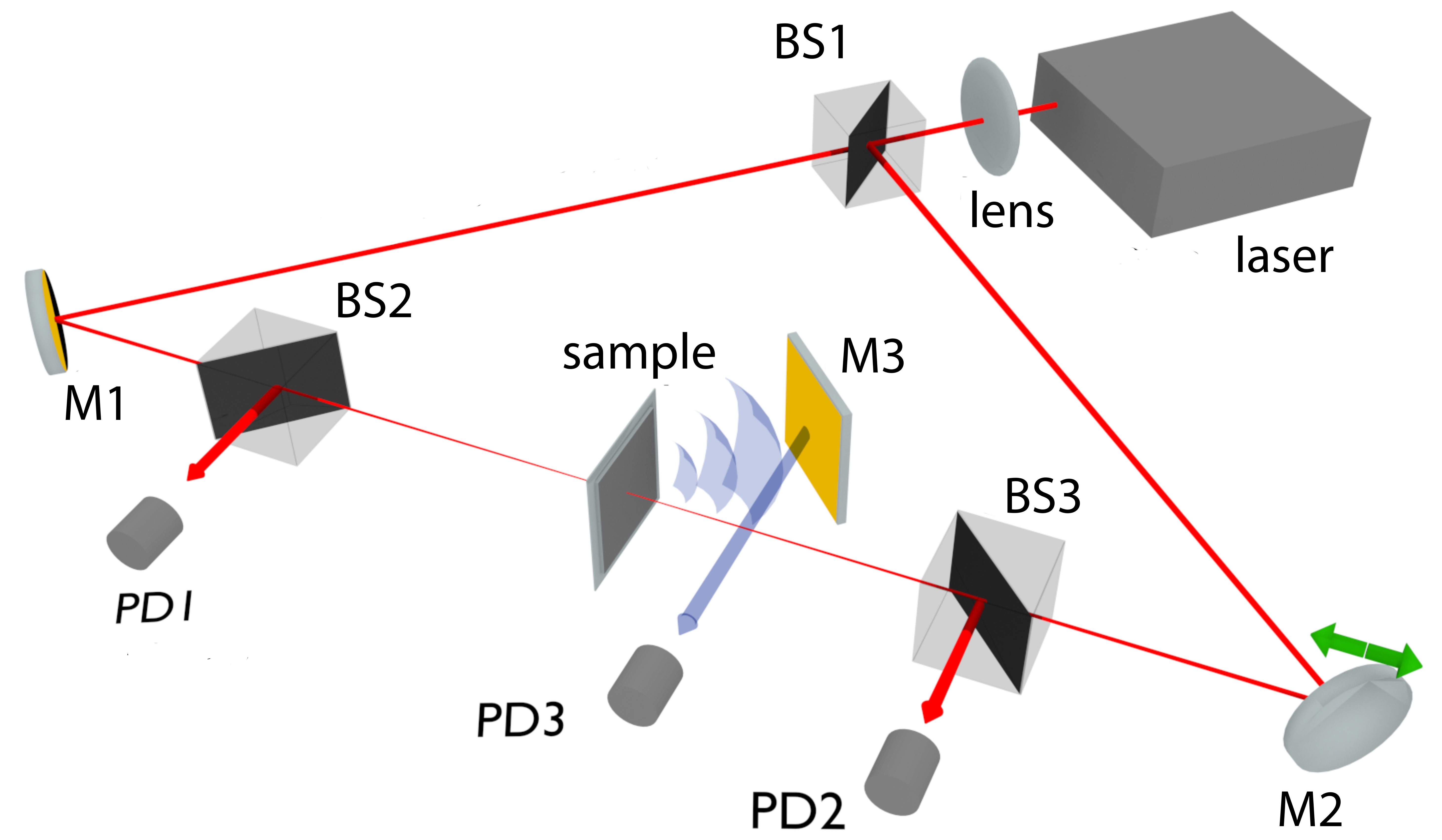}
\caption{A schematic representation of the experimental layout showing the Sagnac interferometer with the graphene sample at the region of interaction of the counter-propagating beams. The photodiodes PD1 and PD2 record the on-axis light energy. A Mirror, M3, collects the scattered light and redirects it to photodiode PD3.}
\label{fig:setup}
\end{figure}
In this letter, we demonstrate for the first time the use of unstructured multilayer graphene films for light-with-light coherent absorption. We also show that this geometry can be used to control, i.e., enhance or suppress, not only absorption but also other optical interactions, such as scattering. Experiments are performed with both continuous-wave and pulsed laser sources attenuated to operate within the linear optics regime. Graphene inherently exhibits the same linear absorption \cite{graphene1, graphene2} over a wide wavelength-range and hence may offer an advantage over using metamaterials or structured films which are fabricated for a certain frequency. We not only demonstrate the modulation of absorption of the optical energy but also that of linear scattering of light by the graphene film by changing the relative phase of the two fields. This shows that along with absorption, other light-matter interactions are also controllable by the light-with-light technique. \\
\indent The experimental layout is shown in Fig(\ref{fig:setup}). The output from a laser source is split into two coherent beams (of similar energies) by a beamsplitter, BS1. The beamsplitter along with the two mirrors M1 and M2, form a Sagnac interferometer. The two counterpropagating beams are loosely focused to the center of the interferometer by a long focal length (f=1 m) lens. The sample is a commercially available multilayer, unstructured, $\sim50$\% absorbing, 30-layer graphene film (Graphene Platform) sandwiched between two fused silica substrates with anti-reflection coating at 532 nm. The sample thickness is $\sim 9$ nm. The graphene sample is placed close to the focal plane of the lens, inside the interferometer and one of the interferometer mirrors, M2, is scanned with a piezoelectrically actuated 1-D translation stage, thus controlling the relative phase of the interfering  beams at the sample position. The reflected and transmitted beams on either side of the sample are monitored via silicon photodiodes PD1 and PD2. The scattered light is collected by a mirror, M3, placed off-axis at an angle of $\sim10^{\circ}$ with respect to the counter-propagating beams and redirected to a third photodiode, PD3.  \\
\indent Fig(\ref{fig:CW}) shows the coherent modulation of the absorption of light at the graphene sample. For this measurement we use a continuous wave (CW), 532 nm laser source with output power of $\sim 500 \mu{W}$ measured at BS1. As the relative phase between the two interacting beams is changed, the graphene layer experiences a sequence of nodes and antinodes in the field pattern and consequently the transmitted energy undergoes a periodic modulation. The modulation of the absorption of light, recorded by PD1 and PD2, is shown by the red squares and blue circles, respectively. The measured transmission is normalised to the total energy input to each arm of the interferometer. We see that the absorption is modulated between 90$\%$ to 10$\%$ - the curves are sinusoidal fits to the data.
The horizontal dashed line shows the single beam absorption of graphene at 532 nm, in the absence of a standing wave, i.e. with one arm of the interferometer blocked. As seen in Fig(\ref{fig:CW}), our sample does not have ideal 50\% absorption, thus explaining the maximum 80\% modulation.\\
We obtain similar results when using femtosecond laser pulses. In this case, the optical pulses are obtained from a Ti:Sapphire laser (Amplitude Technologies) producing 100 fs pulses, centered at 785 nm with a repetition rate of 100 Hz. The energy in each arm of the interferometer was measured to be less than 10 $\mu$J.
Fig(\ref{fig:modulation}) shows the modulation of the transmitted energy as the two counter-propagating  pulses are scanned through each other. The blue curve shows the modulation in the transmitted energy measured at photodiode PD2. The absorption is greatly enhanced or suppressed, with a modulation contrast of $\sim80$\%. The modulation visibility observed here is slightly lesser when compared to the CW case. We attribute this difference to a small chirp in the laser pulse due to propagation through the dispersive optics in the system and also to a slight imbalance in the reflection and transmission coefficients related to the anti-reflection coatings that were not optimised for this wavelength. We simultaneously measured the scattered energy (red solid curve) from the graphene film at PD3 via an off-axis mirror M3, placed at an angle of $\sim 10^{\circ}$ in order to collect the largest portion of scattered light without blocking the on-axis beam. The inset to Fig(\ref{fig:modulation}) shows the full pulse autocorrelation trace obtained by monitoring the scattered energy for a scan over the full pulse length. \\
We note that the on-axis energy and the scattered energy are perfectly out-of-phase. Indeed, at a node of the standing wave there is no net field, hence the on-axis transmission (as discussed above) is maximum. For this same reason, scattering from the film is minimised. At an antinode we should expect, in agreement with the measurements, that the sample experiences maximum field strength so that both absorption and scattering are maximized.  \\
\indent These observations imply that we can effectively control and manipulate absorption and scattering of light using a graphene film placed within a standing wave and thus suggesting an additional route to exploiting graphene, and thin films in general for a range of applications such as optical switching, sensing, etc. \cite{zheludev,integrated}. \\
\begin{figure}[!t]
  \includegraphics[scale=0.3]{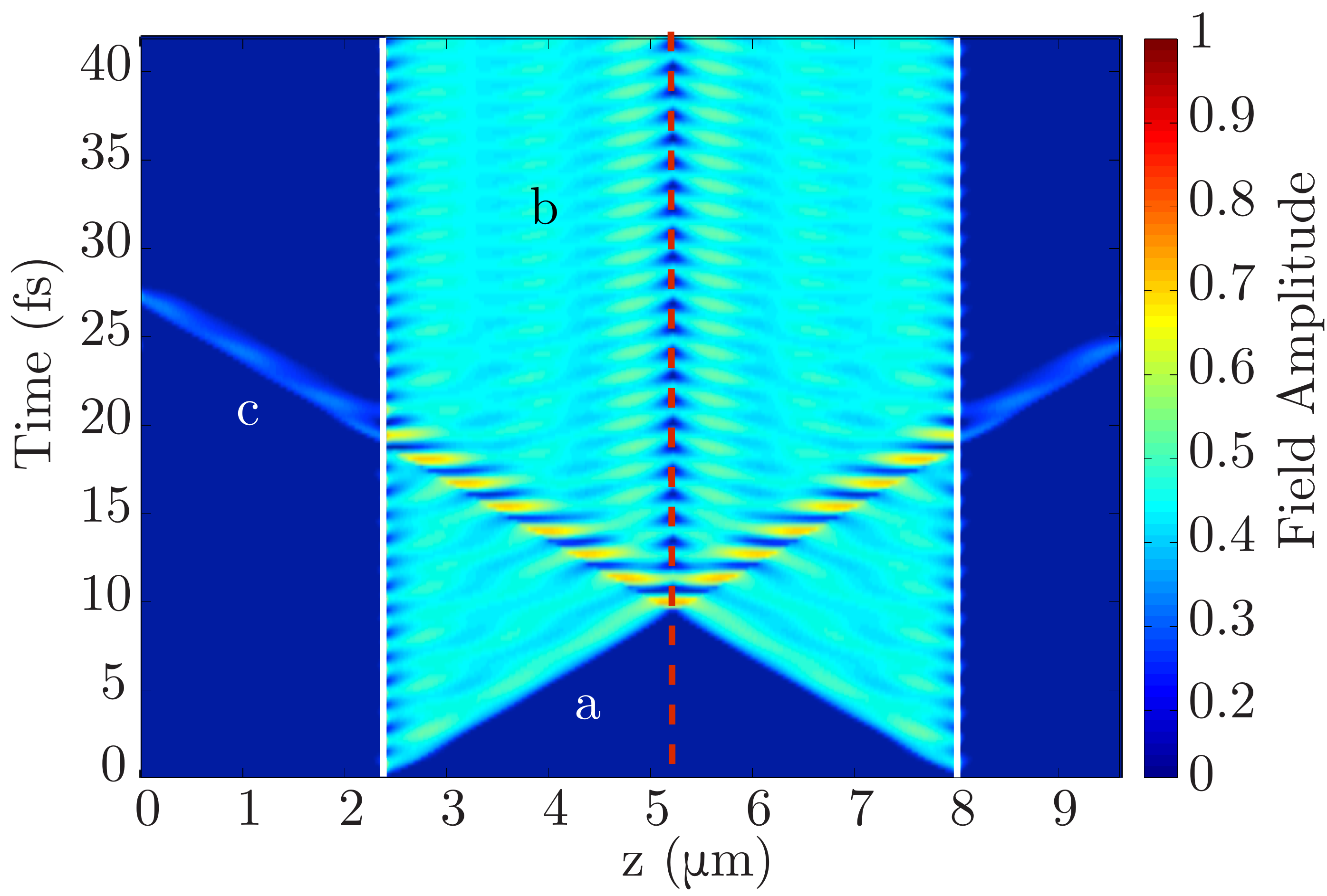}
\caption{z-t plot of the electromagnetic fields evaluated numerically with an FDTD simulation of two continuous wave sources (solid white lines) emitting 785 nm light in opposite directions. ``a'', ``b'' and ``c'' indicate regions of initial build up of the standing wave, transmitted single cycle signal and stationary regime with CPA, respectively }
\label{fig:fdtd}
\end{figure} 
It is possible to create a signal amplitude modulator with the setup described here and integrated versions of such a modulator have also been proposed \cite{zheludev,Zheludev2}. %{\color{blue} In the case of a graphene film, high speed modulation could in principle be achieved by directly modulating the graphene film properties similarly to what has been shown in metamaterial films \cite{Sensale,Yan}.
It is also possible to achieve high speed modulation by directly modifying the thin film properties, as shown in similar studies which use different geometries \cite{Sensale,Yan}. Fig(\ref{fig:fdtd}) shows the propagation evolution ($z-t$ plot) of the two counter-propagating fields evaluated with a numerical finite-difference time-domain (FDTD) simulation.  Two continuous wave sources (positions indicated with solid white lines) emit 785 nm light in opposite directions. The graphene sheet (whose position is indicated with dashed red line) is modelled as a Lorentzian medium \cite{taflove} tuned to transmit and reflect equal amounts ($R = T = 0.25$, such that $\alpha =  50\%$ \cite{abajo}). The mesh size of the simulations performed was 1.25 nm. In region ``a'' the wavefronts, switched on at $t=0$, approach the graphene sheet from either side and start to form a standing wave at around $t\sim10\rightarrow15$ fs. The steady state solution can be seen in region ``b'' ($t>20$ fs), where all light emitted from the sources is absorbed. However, there is a build up time for the standing wave to form during which we see that  $\sim1$ optical cycle of light can leave the set-up, as seen in region ``c''. This indicates that the ultimate limitation on the bandwidth of an optical switch based on CPA in thin films is equal to the frequency of light pulse itself.\\

In conclusion, we have shown that unstructured graphene films with a travelling-wave absorption close to 50$\%$ can be used in light-with-light arrangement so that the total absorption may be continuously controlled between $0\%$ (full transmission) and 100$\%$ (full absorption). Moreover, the same arrangement allows to control other light-matter interaction channels. In the specific case shown here, we are able to modulate, i.e., increase or decrease) the scattering efficiency from the film. We expect this degree of control to extend also to the other effects. For example, graphene is known to exhibit strong nonlinear optical response \cite{nonlinear} and with the experimental configuration used here, there is an interesting possibility of modulating and controlling the nonlinear response of graphene or other promising thin film materials \cite{nonlinear,nonlinear2,nonlinear3}.

\acknowledgements{This research was funded by the European Research Council under the European Union’s Seventh Framework Programme (FP/2007-2013)/ ERC GA 306559 and EPSRC (UK) Grant EP/J00443X/1.}

\end{document}